\documentclass[12pt]{article}

\usepackage[margin=0.7in]{geometry} 
\usepackage{blindtext}
\usepackage{multicol}
\usepackage{setspace}
\usepackage{mathptmx} 
\usepackage{natbib}
\usepackage{graphicx}
\usepackage{amsmath}

\title{Idealized Impacts of Mountainous Terrain on the Energetics of Hurricane Melissa (2025)}
\author{Michael Igbinoba, Princeton University, Department of Computer Science}
\date{December 2025}

\begin{document}
\maketitle

\onehalfspacing 

\begin{multicols}{2}
[
\section{Introduction}
]
The evolution of tropical cyclones interacting with terrain is strongly non-linear, shaped by the specific character of the landmass interaction. Proximity to topography, surface roughness, and the storm’s internal dynamical structure collectively govern how the circulation adjusts as it moves inland. These factors also exert a substantial influence on the distribution and severity of impacts, as the hurricane boundary layer’s transition from smooth ocean surfaces to rough terrestrial terrain varies markedly with the underlying environmental and storm-scale conditions \citep{Chen2025}.

Despite extensive study of storm--rough terrain interactions \citep{Bender1985, Bender1987, Wu2022}, the degree to which stronger tropical cyclones weaken and undergo internal structural changes over rough terrain, relative to their weaker counterparts, remains poorly understood. Hurricane Melissa (2025) was a Category~5 hurricane that made landfall in Jamaica with maximum sustained winds of 160~kt and a minimum central pressure of 892~hPa \citep{NHC2025}. Intense tropical cyclones often exhibit markedly different kinematic and thermodynamic structures, including higher inertial stability, more compact wind fields, shallower boundary layers, and enhanced axisymmetry, all of which may fundamentally modify their response to abrupt increases in surface roughness \citep{Martinez2017, Ren2020, Montgomery2018}. A landfall of this magnitude on terrain with roughness characteristics comparable to Jamaica is unprecedented in the North Atlantic basin, and thus warrants dedicated investigation given the scarcity of studies addressing such extreme events.

In this study, particular attention is given to the decay of internal energetics during Melissa's passage over Jamaica, defined here as the period during which the storm's center remained over the island. Prior studies have examined such energetics observationally by evaluating kinetic energy within a prescribed radius. Following this framework, we analyze the change in kinetic energy within a 200~km radius of the center on a constant geopotential height surface, consistent with the methodology of \citep{Maclay2008}. 

Similarly, integrated kinetic energy (IKE) will be computed within the specified annulus of the real storm before and after land interaction, and these results will be compared to those produced by a novel two-dimensional, Boussinesq, axisymmetric tangential-momentum diffusion model that isolates the effects of vertical turbulent mixing and surface drag from fully three-dimensional asymmetric processes and their associated influences on vortex spin-down. Therefore, the purpose of this study is two-fold: to quantify the kinetic energy lost by Hurricane Melissa during its interaction with Jamaica, and to assess the extent to which an idealized framework can replicate the observed dissipation.

\end{multicols}

\begin{multicols}{2}

\section{Goals and Research Questions}

The dual observational–modeling approach mentioned in the previous section enables the physical mechanisms governing vortex decay over high-roughness terrain to be isolated and evaluated. To achieve this goal, the following research questions are addressed:

\begin{enumerate}
    \item How much IKE did Hurricane Melissa lose as its inner core traversed Jamaica?  
    Quantifying this decay provides a physically interpretable measure of the storm’s weakening that is directly linked to wind-field energetics.

    \item Can a vertically diffusive, axisymmetric tangential-momentum model reproduce the magnitude of the observed IKE loss?
    This evaluates whether vertical turbulent mixing and surface drag alone can account for Melissa’s rapid loss of energy.

    \item What physical processes explain the differences between modeled and observed decay?
    Comparing the two systems reveals the influences of missing physics.

\end{enumerate}

These questions are novel because no prior study has examined energetics of a Category~5 hurricane interacting with high-roughness tropical terrain using an IKE-based framework or compared such observations to a turbulence-only axisymmetric model. Addressing these questions provides deeper insight into the fluid-mechanical controls governing hurricane decay over land.

\end{multicols}

\begin{multicols}{2}
[
\section{Methodology}
]

\subsection{Observational methods}

The observational component of this analysis uses data collected by two NOAA P-3 Hurricane Hunter aircraft during two consecutive missions: 28 October from 1123--1633 UTC and 28--29 October from 2002--0112 UTC. Throughout both missions, the aircraft repeatedly sampled Melissa's eyewall near the 700-hPa level \citep{NHC2024}. Aside from small perturbations caused by intense eyewall turbulence, which are neglected here, the flight-level data correspond to a quasi-constant geopotential height of approximately 3~km. This level therefore serves as the primary analysis height for evaluating the storm's internal energetics. 

To calculate integrated kinetic energy (IKE), we adopt the thin-disk approximation used in \citet{Maclay2008}, treating the storm as a disk of constant depth and radius:
\begin{equation}
\mathrm{IKE_{700}}
= \frac{\rho_0 \, \Delta z}{2}
\int_0^{2\pi} \int_0^{R}
\left( u^2 + v^2 \right) r \, dr \, d\lambda .
\end{equation}
where $\rho_{700}$ denotes the reference air density evaluated on the 700-hPa isobaric surface, $\Delta z$ is the vertical layer thickness, and $u$ and $v$ are the zonal and meridional wind components, respectively. This expression represents the kinetic energy of air parcels azimuthally and radially integrated over the prescribed annulus within a given layer

Assuming axisymmetry ($\partial / \partial \lambda \approx 0$) and noting that the flight-level wind measurements lie near the 700-hPa isobaric surface within a vertical layer of thickness $\Delta z = 1000~\mathrm{m}$, Equation~(1) reduces to
\begin{equation}
\mathrm{IKE}_{700}
= \pi \rho_{0} \Delta z
\int_{0}^{R} v_\lambda^{2}\, r \, dr ,
\end{equation}
where
\begin{equation}
v_\lambda(r) = \sqrt{u^2 + v^2}
\end{equation}
Here, $v_T$ denotes the tangential wind, which is functionally equivalent to the horizontal wind speed computed from the Cartesian wind components. Under the assumption of axisymmetry, this quantity is directly comparable to reconnaissance-observed flight-level wind measurements in a tropical cyclone.

In contrast to \citet{Maclay2008}, this study evaluates the integral within a 100~km annulus rather than 200~km. This narrower radius better isolates the energetics of Melissa’s inner-core region, which is the primary focus. Moreover, the assumption of axisymmetry is generally more appropriate at smaller radii, even for intense hurricanes.

These calculations are applied separately to each flight, after the flight-level wind data are repositioned radially relative to the location of the minimum extrapolated surface pressure. This recentering ensures that the kinetic energy is computed in a storm-centered framework consistent with the axisymmetric assumption.

\subsection{Modeling Methods}

We begin with the axisymmetric, height-dependent radial and tangential momentum equations in cylindrical coordinates $(r,\lambda,z)$. Under the Boussinesq approximation, assuming a constant reference density $\rho_0$, these equations may be written as
\begin{align}
\frac{\partial u_r}{\partial t}
+ u_r \frac{\partial u_r}{\partial r}
+ w \frac{\partial u_r}{\partial z}
- \frac{v_\lambda^2}{r}
&=
- \frac{1}{\rho_0} \frac{\partial p}{\partial r}
+ f v_\lambda
+ F_r,
\label{eq:full_radial}
\\[6pt]
\frac{\partial v_\lambda}{\partial t}
+ u_r \frac{\partial v_\lambda}{\partial r}
+ w \frac{\partial v_\lambda}{\partial z}
+ \frac{u_r v_\lambda}{r}
&=
- f u_r
+ F_\lambda,
\label{eq:full_tangential}
\end{align}
where $u_r(r,z,t)$ and $v_\lambda(r,z,t)$ denote the radial and tangential wind components, respectively, $w(r,z,t)$ is the vertical velocity, $p(r,z,t)$ is pressure, and $f$ is the Coriolis parameter. The terms $F_r$ and $F_\lambda$ represent diffusive and turbulent momentum forcing in the radial and tangential directions.

On horizontal length scales comparable to the radius of gale-force winds, the pressure and geopotential fields are assumed to satisfy hydrostatic balance, such that
\begin{equation}
\frac{\partial p}{\partial z} = - \rho_0 g,
\end{equation}
where $g$ is the acceleration due to gravity.

To obtain a reduced model that evolves only the vertical redistribution of tangential momentum at each radius, several assumptions are introduced. First, the flow is treated as strictly axisymmetric, eliminating all \(\partial/\partial\lambda\) derivatives. Second, radial and vertical advection of tangential momentum are neglected,
\begin{equation}
u_r\frac{\partial v_\lambda}{\partial r} \approx 0,
\qquad
w\frac{\partial v_\lambda}{\partial z} \approx 0,
\end{equation}
consistent with a column model in which each radius is integrated independently. By consequence of no advection, curvature forcing \((u_rv_\lambda/r)\) and Coriolis forcing \((-fu_r)\) in the tangential direction are also omitted, so that the only remaining tendency arises from vertical turbulent flux divergence. Under these assumptions, (\ref{eq:full_tangential}) reduces to
\begin{equation}
\frac{\partial v}{\partial t}
=
F_\lambda.
\label{eq:reduced1}
\end{equation}

The tangential momentum forcing is represented entirely by the vertical divergence of the turbulent stress,
\begin{equation}
F_\lambda
=
\frac{1}{\rho_0}\frac{\partial \tau_\lambda}{\partial z},
\label{eq:tau_def}
\end{equation}
where \(\tau_\lambda\) is the tangential component of the turbulent stress. Combining (\ref{eq:reduced1}) and (\ref{eq:tau_def}) yields the one-dimensional governing equation for tangential wind:
\begin{equation}
\frac{\partial v_\lambda}{\partial t}(z,t)
=
\frac{1}{\rho_0}\frac{\partial \tau_\lambda}{\partial z}(z,t).
\label{eq:dvdt_final}
\end{equation}

The vertical stress is represented using a nonlinear drag law at the lower boundary associated with high Reynold number flows and turbulent diffusion aloft:
\begin{equation}
\tau_\lambda(z,t)
=
\begin{cases}
\rho_0 C_D\, v_\lambda(t)\,|v_\lambda(t)|, & z=0, \\[4pt]
\rho_0 K(z)\,\dfrac{\partial v_\lambda}{\partial z}(z,t), & 0<z\le H_{\mathrm{top}},
\end{cases}
\label{eq:stress_def}
\end{equation}
where \(C_D\) is the bulk surface drag coefficient, \(v_s(t)\) is the near-surface tangential wind, and \(K(z)\) is a height-dependent eddy viscosity. The turbulent viscosity decreases exponentially with height according to
\begin{equation}
K(z)
=
K_{T}\,\exp\!\left(-\frac{z}{z_{\mathrm{BL}}}\right),
\label{eq:K_profile}
\end{equation}
where \(K_{T}\) is a characteristic boundary-layer eddy viscosity and \(z_{\mathrm{BL}}\) is an e-folding depth set at 1~km, following results from GPS dropwindsonde observed boundary layer depths in stronger hurricanes \citep{Zhang2012}.  

Substituting (\ref{eq:stress_def}) into (\ref{eq:dvdt_final}) yields the governing equation for the evolution of tangential wind:
\begin{equation}
\frac{\partial v_\lambda}{\partial t}(z,t)
=
\frac{\partial}{\partial z}
\left[
K(z)\,\frac{\partial v_\lambda}{\partial z}(z,t)
\right],
\qquad 0<z\le H_{\mathrm{top}},
\label{eq:final_column_eq}
\end{equation}
with the nonlinear boundary condition
\begin{equation}
\left. K(z)\,\frac{\partial v_\lambda}{\partial z}\right|_{z=0}
=
C_D\, v_\lambda(t)\,|v_\lambda(t)|.
\label{eq:bc_surface}
\end{equation}

Equations (\ref{eq:final_column_eq})–(\ref{eq:bc_surface}) describe the evolution of tangential wind at each radius solely through vertical turbulent mixing and surface drag. This forms the basis of the axisymmetric diffusion model used in this study.

The vertical domain extends from the surface to \(H_{\mathrm{top}}=20\) km and is discretized into 100 uniform layers of thickness
\begin{equation}
h = \frac{H_{\mathrm{top}}}{100},
\qquad
z_i = i h,\quad i=0,\dots,100,
\end{equation}
with layer-mean tangential wind
\begin{equation}
v_k(t)
=
\frac{1}{h} \int_{z_{k-1}}^{z_k} v(z,t)\,dz,
\qquad k=1,\dots,100.
\end{equation}
Applying the operator $(1/h)\int_{z_{k-1}}^{z_k}(\cdot)\,dz$ to (\ref{eq:final_column_eq}) yields the layer-mean tendency equation in flux-divergence form,
\begin{equation}
\frac{\partial v_k}{\partial t}
=
\frac{1}{\rho_0 h}
\left[
\tau_\lambda\!\left(z_{k+\frac12},t\right)
-
\tau_\lambda\!\left(z_{k-\frac12},t\right)
\right],
\label{eq:general_discrete}
\end{equation}
where $v_k(t)$ is the layer-mean tangential wind over $z\in[z_{k-1},z_k]$, and the turbulent stresses are evaluated at the layer interfaces
$z_{k\pm \frac12}$.

The surface stress is obtained from (\ref{eq:stress_def}) at the lower boundary and is approximated using the lowest-layer mean wind,
\begin{equation}
\tau_\lambda\!\left(z_{\frac12},t\right)
=
\rho_0 C_D\, v_1(t)\,|v_1(t)| .
\label{eq:surf_stress}
\end{equation}
For interior interfaces, the diffusive stress is evaluated using a second-order, interface-centered finite-difference approximation,
\begin{equation}
\tau_\lambda\!\left(z_{i+\frac12},t\right)
=
\rho_0 K_i^\ast
\frac{v_{i+1}(t)-v_i(t)}{h},
\qquad i=1,\dots,99,
\label{eq:int_stress}
\end{equation}
where $K_i^\ast$ is the interface eddy viscosity formed by arithmetic averaging between adjacent layer-center values. A stress-free upper boundary,
\begin{equation}
\tau_\lambda\!\left(z_{100+\frac12},t\right)=0,
\label{eq:top_stressfree}
\end{equation}
closes the system.

Substituting (\ref{eq:surf_stress})--(\ref{eq:top_stressfree}) into (\ref{eq:general_discrete}) and canceling the constant density yields the discrete evolution equations describing momentum diffusion and surface drag. The lowest layer satisfies
\begin{equation}
\frac{\partial v_1}{\partial t}
=
\frac{K_1^\ast}{h^2}\,(v_2-v_1)
-
\frac{C_D}{h}\,v_1 |v_1|,
\label{eq:v1_eq}
\end{equation}
which includes both turbulent diffusion and nonlinear drag. For interior layers $k=2,\dots,99$,
\begin{equation}
\frac{\partial v_k}{\partial t}
=
\frac{1}{h^2}
\left[
K_k^\ast (v_{k+1}-v_k)
-
K_{k-1}^\ast (v_k - v_{k-1})
\right],
\label{eq:interior_eq}
\end{equation}
representing vertically varying diffusion. The top layer satisfies
\begin{equation}
\frac{\partial v_{100}}{\partial t}
=
\frac{K_{99}^\ast}{h^2}\,(v_{99}-v_{100}),
\label{eq:top_eq}
\end{equation}
consistent with the imposed zero-stress boundary condition.

A diagnostic geopotential height field is obtained via gradient-wind balance. Neglecting radial acceleration and friction in the radial momentum equation, the radial pressure gradient at each height satisfies
\begin{equation}
\frac{1}{\rho_0}\frac{\partial p}{\partial r}
=
\frac{v_\lambda^2}{r} + f v_\lambda.
\end{equation}
Under hydrostatic balance, \(\partial p/\partial z = -\rho_0 g\), the radial gradient of geopotential becomes
\begin{equation}
\frac{\partial \Phi}{\partial r}
=
\frac{v^2(r,z)}{r} + f v(r,z),
\end{equation}
which is integrated radially outward to an environmental radius to obtain \(\Phi(r,z)\). The geopotential height perturbation is then \(Z(r,z)=\Phi(r,z)/g\), providing a dynamical measure of the warm core consistent with the evolving tangential wind.

The drag coefficient \(C_D\) is selected based on roughness lengths appropriate for mountainous terrain. Using the logarithmic boundary-layer relationship
\begin{equation}
C_D
\approx
\left[
\frac{\kappa}{\ln(\frac{z_{\mathrm{ref}}}{z_0^{\mathrm{eff}}})}
\right]^2,
\end{equation}
and adopting \(z_0^{\mathrm{eff}}\) of order meters observed in similar terrain \citep{Han2014}, the resulting drag is of order \(10^{-2}\), and a representative value \(C_D=2\times10^{-2}\) is used. The eddy viscosity \(K_{T}\) is chosen within the typical range inferred for strongly mixed tropical-cyclone boundary layers (O[10\(^2\) m\(^2\) s\(^{-1}\)]), then decays exponentially following (\ref{eq:K_profile}). Dropsonde-based studies within the lower few-hundred meters of the hurricane boundary layer have found values of $K_T$ around $100~\mathrm{m^2\,s^{-1}}$, but given the increased shallowness of Melissa's boundary layer and lack of observations near Melissa's intensity, this value could be even higher within the jet that resides at the lowest few model layers \citep{Zhang2012}.

At each time step, the model computes integrated kinetic energy at $z = 3$ km using the same expression applied to the observational data:
\begin{equation}
\mathrm{IKE}_{3\mathrm{km}}
=
\pi \rho_0 \Delta z
\int_0^{R_{\max}}
v_\lambda^2(r,z,t) \, r\, dr,
\end{equation}
facilitating direct comparison between the simulated energetics decay and that observed during Hurricane Melissa’s land interaction. Model parameters are in Table~1.

\section{Results}

\subsection{Observational Results}

\begin{figure}[htbp]
\centering
\includegraphics[width=0.9\textwidth]{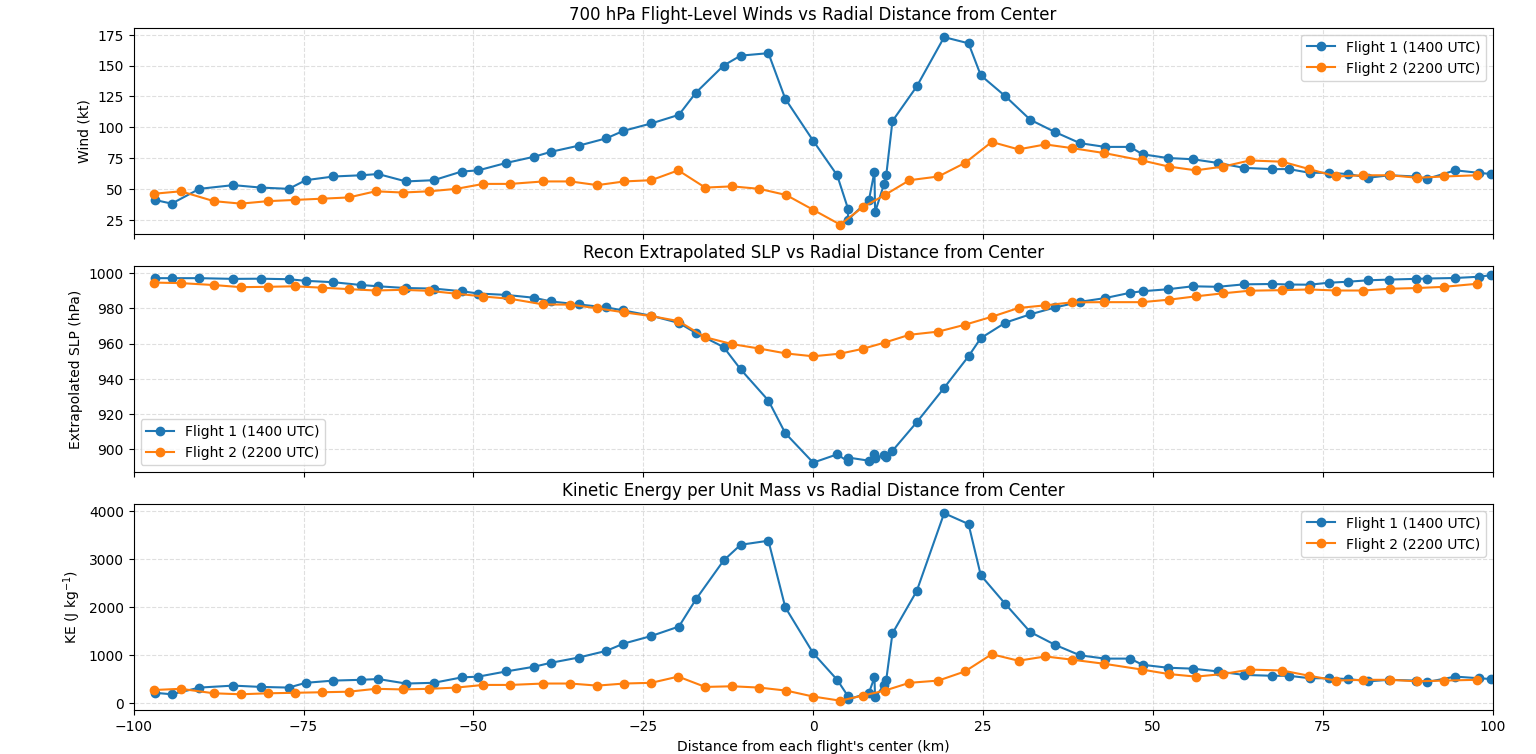}
\caption{Observed azimuthally averaged tangential wind, pressure, and IKE diagnostics from the first and second reconnaissance flights used in this study.}
\label{fig:fig1}
\end{figure}

Utilizing the IKE formulation described in the observational methods section, a direct comparison between the first and second reconnaissance flights was performed. As shown in Figure~\ref{fig:fig1}, the storm experienced a drastic spin-down during its interaction with Jamaica. The observed flight-level maximum tangential wind decreased by approximately 48\%, from 173~kt to 90~kt, and the radius of maximum winds expanded outward by about 11~km (from 20~km to 31~km), indicating a pronounced weakening and broadening of the circulation.

In addition, the minimum surface pressure rose by 58~hPa, increasing from 892~hPa to 950~hPa, a rapid fill-in rate comparable to that documented for Hurricane Patricia (2015) \citep{Rogers2017}. Most notably, the integrated kinetic energy exhibited a substantial decline, decreasing by 41\% from $1.9\times10^{16}$~J to $1.1\times10^{16}$~J.

\subsection{Model Results}

\begin{figure}[htbp]
\centering
\includegraphics[width=0.9\textwidth]{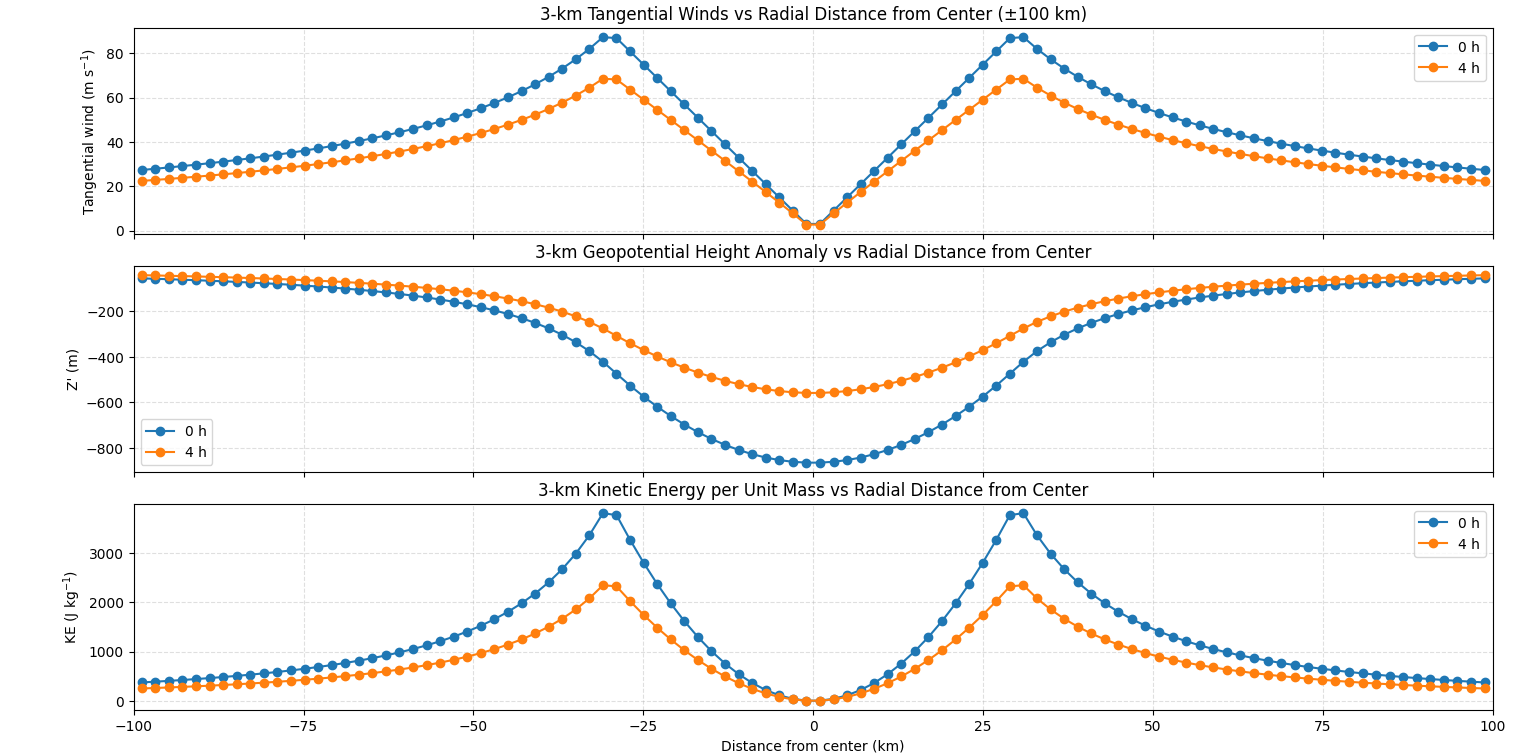}
\caption{3-km tangential wind, kinetic energy per unit mass, and geopotential height from the axisymmetric diffusion model at initialization and after 4~h of integration.}
\label{fig:fig2}
\end{figure}

\begin{figure}[htbp]
\centering
\includegraphics[width=0.9\textwidth]{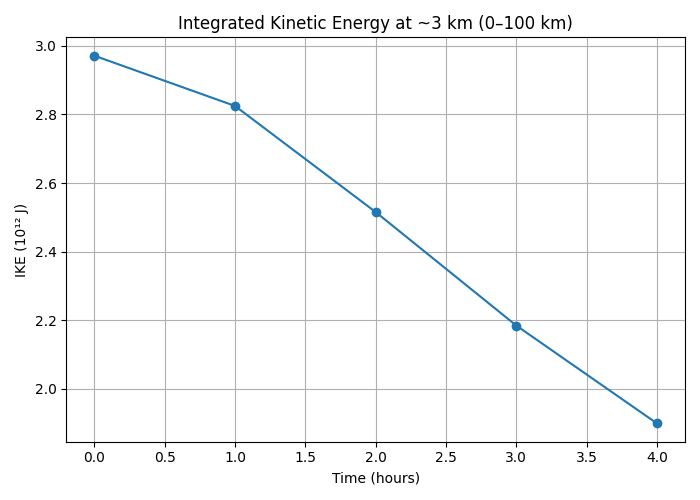}
\caption{Time evolution of integrated kinetic energy (IKE) at 3~km within 100~km of the center from the axisymmetric diffusion model.}
\label{fig:fig3}
\end{figure}

Using the same diagnostic framework applied to the observational data, results from the axisymmetric diffusion model were analyzed at the initial time and after four hours of integration. As shown in Figure~\ref{fig:fig2}, the model also produced rapid weakening. The maximum tangential wind at the 3~km level decreased from 175~kt to 134~kt, a reduction of approximately 23\%, while the central geopotential height at 3~km rose by roughly 300~m, consistent with a substantial decrease in vortex strength. The integrated kinetic energy likewise declined, decreasing by 36\% from $2.9\times10^{16}$~J to $1.8\times10^{16}$~J. Furthermore, a linear trend in IKE could be identified from Figure~\ref{fig:fig3}, with decrease being roughly constant from intialization; however, this is likely a manifestation of the diffusion operator progressively damping higher-order vertical modes, leaving behind increasingly slow-decaying structures whose collective decay rate appears nearly linear over short time intervals.

\subsection{Comparison}
Using both Figures~\ref{fig:fig1} and \ref{fig:fig2}, both the reconnaissance data and the axisymmetric diffusion model show substantial reductions in tangential wind, outward displacement of the wind maximum (either explicitly observed or implicitly indicated by broadening of the model geopotential height field), an increase in central geopotential height or pressure, and a large decline in IKE over the analysis period. The fact that the simple diffusion model captures these leading-order features demonstrates that vertical turbulent mixing and enhanced surface drag alone can account for much of Melissa’s observed rapid decay.

Quantitatively, however, the observed weakening was considerably more dramatic than the modeled weakening. The real storm experienced a 48\% reduction in peak tangential wind, compared to only 23\% in the model, and a 41\% decrease in IKE versus a 36\% decrease in the simulated vortex. The observations also show a more pronounced expansion of the radius of maximum winds and a substantially larger increase in central pressure than the model suggests.

Despite these limitations, the similarity in the magnitude and direction of the modeled and observed energy loss indicates that extreme surface roughness over Jamaica was a primary driver of Melissa's rapid spin-down. The model’s ability to reproduce a substantial fraction of the observed decay, using only friction and vertical mixing, suggests that additional processes in the real storm likely acted to amplify the underlying frictionally driven decay mechanism.

\subsection{Discussion}
Hurricane Melissa's extreme degradation over land, despite otherwise favorable dynamic and thermodynamic background conditions, likely reflects the extraordinary roughness of the terrain it encountered. With an estimated effective roughness length on the order of meters, consistent with values reported for similar environments \citep{Han2014}, frictional dissipation of kinetic energy is greatly amplified due to the large surface drag imposed on the near-surface flow. This enhanced drag directly weakens both the primary circulation (through rapid reduction of tangential wind) and the secondary circulation (through reduced inflow and vertical mass transport).

Simultaneously, the transition from the warm ocean surface to the cooler, drier land surface sharply suppresses surface enthalpy fluxes, cutting off the storm’s thermodynamic energy supply. Orographically forced downsloping flow further entrains lower-$\Theta_e$ air into the boundary layer, stabilizing it and diminishing its ability to saturate and support deep convection \citep{Ahern2021}. The combination of boundary-layer stabilization and enhanced frictional mixing both deepens the boundary layer and weakens the inflow-driven secondary circulation. This reduction in secondary circulation limits eyewall forcing, weakens the warm core aloft, and contributes to a rise in central pressure through Sawyer–Eliassen responses and asymmetric dynamical effects \citep{Emanuel1986}. Together, these interacting negative feedbacks accelerate the spin-down of the vortex as it traverses Jamaica.

The performance of the axisymmetric diffusion model, despite its simplicity and omission of key dynamical processes such as asymmetric eddies, terrain-induced flow distortion, radial and vertical advection, and thermodynamic feedbacks, further suggests that the extreme surface roughness of the Jamaican landmass was a first-order mechanism governing Melissa’s weakening. That such a reduced-physics model reproduces a substantial fraction of the observed IKE loss indicates that enhanced friction and vertical turbulent mixing alone can account for much of the storm’s energetics decay, with additional processes likely acting to further accelerate the weakening observed in the reconnaissance data.

\section{Conclusion and future work}
This study examined the decay of Hurricane Melissa’s inner-core energetics as the storm traversed the mountainous terrain of Jamaica and evaluated the ability of a simplified axisymmetric diffusion model to reproduce the observed weakening. Using reconnaissance flight data, a substantial reduction in integrated kinetic energy (IKE) was documented, with a 41\% loss occurring over the 4~h period during which Melissa’s center was over land. This decline coincided with marked structural degradation, including a 48\% reduction in maximum tangential winds, outward migration of the radius of maximum winds, and a 58~hPa rise in central pressure.

Results from the axisymmetric diffusion model demonstrated that vertical turbulent mixing and strong surface drag alone could account for a significant portion of the observed decay. Although the model underestimates the total weakening compared to the flight data, it nevertheless reproduces a substantial fraction of the IKE loss, indicating that enhanced frictional dissipation over Jamaica’s high-roughness terrain was a first-order mechanism governing Melissa’s rapid spin-down. The discrepancy between the model and observations further highlights the contributions of physical processes absent from the reduced framework, such as asymmetric eddies, terrain-induced flow distortion, radial momentum advection, eyewall collapse, and thermodynamic influences on secondary circulation strength.

A broader implication of this work is that even highly idealized models can capture the leading-order energetics of tropical cyclone decay over rough terrain when friction and vertical mixing dominate the response. This underscores the value of simplified fluid-mechanical models in isolating fundamental decay mechanisms, while also illustrating the limitations of axisymmetric frameworks in representing terrain-induced asymmetries and internal dynamical turbulence.

Future work should incorporate additional physical processes to improve realism, including radial momentum transport, asymmetric forcing from complex terrain, and interactive surface fluxes that govern thermodynamic feedbacks. These enhancements can be achieved by employing a fully three-dimensional, full-physics modeling framework such as the Weather Research and Forecasting (WRF) model \citep{Skamarock2019}, in which the representation of boundary-layer dynamics, radiative and moist processes, and topographic influences is substantially more complete. Incorporating observed and modeled initial and boundary conditions, particularly those derived from global or regional analyses, would further ground the simulations in realistic synoptic environments, enabling the study of how large-scale features modulate decay as storms encounter rough terrain.

Extending this analysis to other intense storms that have interacted with high-roughness landmasses would also help evaluate the generality of the mechanisms identified here and within more sophisticated numerical systems. Finally, deeper coupling of a less-idealized modeling framework with high-resolution simulations and the broader suite of reconnaissance observations, many of which could not be utilized in this study due to compatibility or time limitations, would allow for a more comprehensive assessment of how frictional, thermodynamic, and topographic effects jointly influence vortex decay in Hurricane Melissa and similar environmental settings.

\end{multicols}


\begin{table}[ht!]
\centering
\footnotesize
\caption{Model parameters used in the axisymmetric tangential-momentum diffusion simulation.}
\label{tab:model_params}
\begin{tabular}{p{4cm} p{2.5cm} p{1.5cm}}
\textbf{Parameter} & \textbf{Value / Symbol} & \textbf{Units} \\
\hline
Reference density & $\rho_{0} = 0.9$ & kg m$^{-3}$ \\
Gravity & $g = 9.81$ & m s$^{-2}$ \\
Coriolis parameter & $f = 4.5\times10^{-5}$ & s$^{-1}$ \\
Surface drag coefficient & $C_D = 2.0\times10^{-2}$ & -- \\
0th layer eddy viscosity & $K_{T0} = 300$ & m$^2$ s$^{-1}$ \\
BL height & $z_{BL} = 1000$ & m \\
Vertical layers & $N = 100$ & -- \\
Model top & $H_{\text{top}}=20$ & km \\
Radial extent & $r_{\max}=500$ & km \\
Radial points & $n_r=251$ & -- \\
Radius of max wind & $R_m=30$ & km \\
Max tangential wind & $V_{\max}=175$ & kt \\
Peak height of profile & $z_{\text{peak}} = 1$ & km \\
Gaussian width & $\sigma_z = 5$ & km \\
Time step & $\Delta t = 10$ & s \\
Simulation duration & 4 & hours \\
\hline
\end{tabular}
\end{table}


\end{document}